# Momentum Dynamics in Competitive Sports: A Multi-Model Analysis Using TOPSIS and Logistic Regression

Mingpu Ma[*]

* College of Management and Economics, Tianjin University, Tianjin 300072, China
E-mail: m125_0409@tju.edu.cn

**Summary:** Currently, sports competitions are being more emphasized. Sports not only can effectively enhance people's fitness, but also is the intersection of physiology, medicine, statistics and other disciplines. In sports competitions, there are often many extraordinary battles. Some athletes who are at a disadvantage can turn the tables. This is due to the concept of momentum. In order to investigate the "momentum" in the game, we developed **Model 1: TOPSIS Model** and **Model 2: 0-1 Logistic Regression Model**. These two models are tightly coupled to explain the generation of momentum and to make predictions about the situation.

To investigate match flow modelling, we developed a **TOPSIS Evaluation Model** to capture the dynamics between two tennis players, using data from the 2023 Wimbledon Open. By visualizing this data, we analyzed player performance and advantages at various points in the match, illustrating how "momentum" manifests in sports.

To demonstrate that "momentum" influences sports outcomes, we constructed a **0-1 Logistic Regression Model**, using player performance fluctuations as the independent variable and success scores as the dependent variable. The model's high predictive accuracy confirms that these performance fluctuations and successive successes are not random, underscoring the significant role of "momentum" in sports.

Next, we use **logistic regression** to determine the indicators that can prove the turn of the game situation. We can see that there is a regularity in the momentum change indicators of the players in the key games, which can be used to determine whether the players in the key games are in an advantageous position in the game. Subsequently, we used **Spearman's Correlation Coefficient Analysis** to correlate the momentum change indicators, and then we obtained eight variables with the strongest correlation. Since the correlation coefficients of these eight variables were not very different, we applied **Principal Component Analysis** to find the most relevant factors, according to which we gave player recommendations.

Finally, we combined the **TOPSIS Model** with **0-1 Logistic Regression Model** to test the accuracy of model with match data. The test results show that our model is able to explain the conditions that occur during the match and with a high degree of accuracy. We then discussed factors that could be incorporated into model, and the generalizability of the model is explored.

In the end, we wrote a memo to coaches to explain our results and make suggestions. The strengths and weaknesses of our model and its future improvement are also analyzed.

# Contents

# 1 Introduction

## 1.1 Background

Momentum is often considered a physics-related concept; in physics momentum is a physical quantity related to the mass and velocity of an object and refers to the effect of a moving object. In some literature, momentum is defined as "the momentum or force obtained by motion or a series of motions". For example, in the 2023 Wimbledon men's final, in which Carlos Alcaraz defeated Novak Djokovic, the situation changed unimaginably in each game from the first to the fifth set. Unexpected fluctuations of a seemingly dominant player in each set are the concept of "momentum" as defined in the literature.

Therefore, this paper considers momentum to be the direction of the flow of a situation in certain rounds of a match, as well as the force that causes the situation to change. In many matches, players have the feeling that "the more they play, the better they feel", which is the effect of momentum. Although momentum is difficult to measure in a game, studying it is important for us to study the development of the current situation and allows us to make better plans for the future.

Considering the background information and restricted conditions identified in the problem statement, we need to solve the following problems:

- We need to develop a model that properly captures the "momentum" of the game and apply it to multiple games. Our model will determine the performance advantage of a player at a given point in the game.

- We will model that fluctuations in player performance and successive occurrences of success are not random, but are due to "momentum".

- We will use the given data to determine the specific indicators that can shift the game from one player to another, thus guiding the order in which the players play in a sports game.

- We will investigate the accuracy of our model, and we will explore the factors that improve the model in this paper. Finally, we will disseminate our model so that it can be used in other games or sports.

## 1.2 Literature Review

There are fewer studies in the literature about the "momentum" of the race, but the research started earlier. In 1999, Tian et al. studied how basketball players' mental training can better improve their game momentum [1], which is important for the subsequent research on "momentum", and in 2002, Fu et al. analyzed the specific clinical factors of basketball games, and analyzed the game momentum of the game [2], which is important for the subsequent research on "momentum". "In 2008, Wang et al. studied the substitution tactics in basketball games, and proposed that substitution tactics can play the role of a surprise attacker or even reverse the game [3], which formally put forward the concept of "momentum" of the game. In 2017, Li et al. extended the concept of game momentum to the field of table tennis, and analyzed the tactics of table tennis [4]. Tan et al. specifically analyzed the tactics of table tennis, and used the concept of momentum to analyze different moments of the game [5]. In 2022, You et al. studied the pause timing of table tennis matches, and analyzed the pause moments of different nationalities in different matches, and concluded the real existence and differences of the game rhythm and "momentum" [6]. 2023, Sun et al. applied the game "momentum" to analyze the game rhythm and "momentum" [7]. "Momentum" to construct the volleyball coaches' in-game tactical decision-making system, so that this concept can guide the tactical use in specific games, which has played a good effect.

## 1.3 Our Work

By exploring the background of the problem and reanalyzing it, we have organized the ideas of this paper in the form of a fishbone diagram as follows.

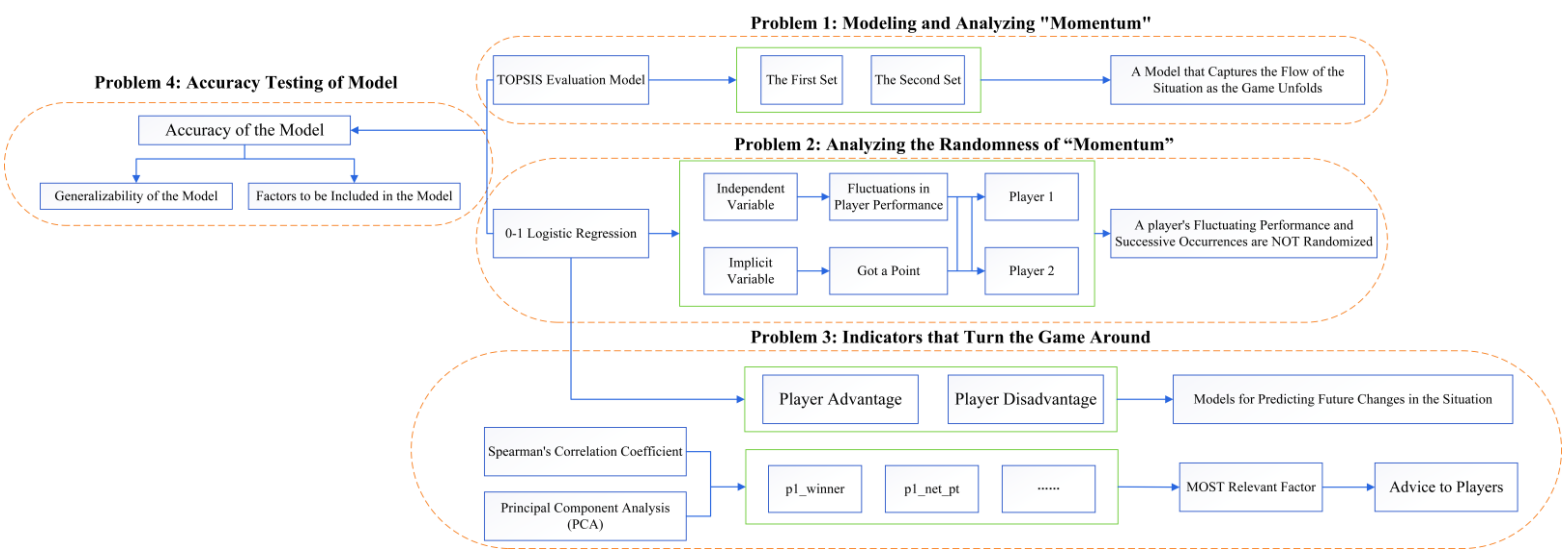

Figure 1: The Flow Chart of Our Work

# 2 Assumptions and Glossaries

To simplify the above problem, we follow the following assumption in the problem-solving process.

**Assumption1**: For the assessment of which player's performance is superior and the situation is reversed, we consider that the player's physical and psychological qualities have no influence on the game.

**Justification**: When assessing the performance of a player, each player's physical and psychological qualities will have some impact on the performance during the game, but in order to simplify the model, we do not consider the impact of the player's physical qualities, psychological qualities on the performance advantage of the game as well as the reversal of the situation.

**Assumption2**: For assessing players' performance advantage and situation reversal, we consider that the venue has no effect.

**Justification**: Different venues have different effects on player performance, so it is important to consider the factors of the playing venue as well. However, in order to simplify the model, we do not consider the effect of the venue on assessing a player's performance advantage and situations where the situation is reversed.

**Assumption3**: For assessing which player performs better, we consider that the player's playing style and bat speed have no effect on the match.

**Justification**: Each player's batting style is different and has a certain impact on the change of situation and advantage of the game. In order to simplify the model, we do not consider the impact of the player's batting style and batting speed on the player's performance advantage when using the TOPSIS Evaluation Model.

**Assumption4**: For the evaluation of which player's performance advantage and the reversal of the situation, we consider that the behavior of a player exploiting an opponent's weakness has no impact on the game.

**Justification**: Being able to recognize and exploit an opponent's weaknesses may have a critical impact on changing the situation. However, in order to simplify the model, we do not consider the impact of identifying and exploiting an opponent's weakness on the match.

The necessary mathematical notations used in this paper are listed in Table 1.

**Table 1: Glossaries used in this paper**

| Variables | Explanation | Example |
|---|---|---|
| $match\_id$ | match identification | 2023-wimbledon-1701 ("7" is the round, and "01" the match number in that round) |
| $player1$ | first and last name of the first player | Carlos Alcaraz |
| $player2$ | first and last name of the second player | Novak Djokovic |
| $elapsed\_time$ | time elapsed since start of first point to start of current point (H:MM:SS) | 0:10:27 |
| $set\_no$ | set number in match | 1, 2, 3, 4, or 5 |
| $game\_no$ | game number in set | 1, 2, ...,7 |
| $point\_no$ | point number in game | 1, 2, 3... etc. |
| $p1\_sets$ | sets won by player 1 | 0, 1, or 2 |
| $p2\_sets$ | sets won by player 2 | 0, 1, or 2 |
| $p1\_games$ | games won by player 1 in current set | 0, 1,...,6 |
| $p2\_games$ | games won by player 2 in current set | 0, 1,...,6 |
| $p1\_score$ | player 1's score within current game | 0 (love), 15, 30, 40, AD (advantage) |
| $p2\_score$ | player 2's score within current game | 0 (love), 15, 30, 40, AD (advantage) |
| $server$ | server of the point | 1/2: player 1/2 |

| Variables | Explanation | Example |
|---|---|---|
| serve_no | first or second serve | 1/2: first/second serve |
| point_victor | winner of the point | 1/2: if player 1/2 wins |
| p1_points_won | number of points won by player 1 in match | 0, 1, 2... etc. |
| p2_points_won | number of points won by player 2 in match | 0, 1, 2... etc. |
| game_victor | a player won a game this point | 0: no one, 1/2: player 1/2 |
| set_victor | a player won a set this point | 0: no one, 1/2: player 1/2 |
| p1_ace | player 1 hit an untouchable winning serve | 0 or 1 |
| p2_ace | player 2 hit an untouchable winning serve | 0 or 1 |
| p1_winner | player 1 hit an untouchable winning shot | 0 or 1 |
| p2_winner | player 2 hit an untouchable winning shot | 0 or 1 |
| winner_shot_type | category of untouchable shot | F: Forehand, B: Backhand |
| p1_double_fault | player 1 missed both serves and lost the point | 0 or 1 |
| p2_double_fault | player 2 missed both serves and lost the point | 0 or 1 |
| p1_unf_err | player 1 made an unforced error | 0 or 1 |
| p2_unf_err | player 2 made an unforced error | 0 or 1 |
| p1_net_pt | player 1 made it to the net | 0 or 1 |
| p2_net_pt | player 2 made it to the net | 0 or 1 |
| p1_net_pt_won | player 1 won the point while at the net | 0 or 1 |
| p2_net_pt_won | player 2 won the point while at the net | 0 or 1 |
| p1_break_pt | player 1 has an opportunity to win a game player 2 is serving | 0 or 1 |
| p2_break_pt | player 2 has an opportunity to win a game player 1 is serving | 0 or 1 |
| p1_break_pt_won | player 1 won the game player 2 is serving | 0 or 1 |
| p2_break_pt_won | player 2 won the game player 1 is serving | 0 or 1 |
| p1_break_pt_missed | player 1 missed an opportunity to win a game player 2 is serving | 0 or 1 |
| p2_break_pt_missed | player 2 missed an opportunity to win a game player 1 is serving | 0 or 1 |
| p1_distance_run | player 1's distance ran during point (meters) | 5.376, 21.384, etc. |
| p2_distance_run | player 2's distance ran during point (meters) | 6.485, 12.473, etc. |
| rally_count | number of shots during the point | 1, 2, 4, etc. (includes serve) |
| speed_mph | speed of serve (miles per hour; mph) | 81, 124, etc. |
| serve_width | direction of serve | B: Body, BC: Body/Center, BW: Body/Wide, C: Center, |

| Variables | Explanation | Example |
|---|---|---|
| *serve_depth* | depth of serve | W: Wide CTL: Close To Line, NCTL: Not Close To Line |
| *return_depth* | depth of return | D: Deep, ND: Not Deep |

# 3 Modeling and Analyzing "Momentum"

In that section, we used the TOPIS evaluation method to select the number of discs and the number of innings as evaluation indexes, and we also constructed a visualization image to describe the match trend.

## 3.1 Selection of TOPSIS Model

The TOPSIS (Technique for Order of Preference by Similarity to Ideal Solution) evaluation method is a multi-attribute decision analysis technique used to assess the relative merits of multiple alternative solutions. It is based on the concept of distance and determines the relative superiority of alternative solutions by calculating their distances from the ideal solution. The specific steps of the solution are as follows:

**Step1**: First, identify the various attribute indicators for evaluating the solutions, denoted as $A_1, A_2, ..., A_m$, where $m$ represents the number of attribute indicators.

**Step2**: Assign weights to each attribute indicator to represent their importance in the decision-making process. Let the weight vector be $W = (w_1, w_2, ..., w_m)$, where $w_i$ denotes the weight of the $i$-th attribute indicator, satisfying $\sum_{i=1}^{m} w_i = 1$.

**Step3**: Arrange the attribute indicator values of each alternative solution in matrix form, creating a decision matrix $X$. Each row of the matrix represents the attribute indicator values of a specific alternative solution, and each column represents the values of a particular attribute indicator. The decision matrix $X$ has dimensions $n \times m$, where $n$ represents the number of alternative solutions.

**Step4**: Normalize the decision matrix: Normalize each column of the decision matrix $X$ to convert the attribute indicator values into dimensionless relative values. Normalization can be performed using the following formula:

$$x_{ij}^* = \frac{x_{ij}}{\sqrt{\sum_{i=1}^{n} x_{ij}^2}} \tag{1}$$

Here, $x_{ij}^*$ represents the normalized value of the attribute indicator.

**Step5**: Based on the decision-maker's preferences, determine the ideal solution vector

$A^* = (a_1^*, a_2^*, ..., a_m^*)$ and the negative ideal solution vector $A^- = (a_1^-, a_2^-, ..., a_m^-)$. Each element $a_i^*$ of the ideal solution vector represents the best value for the $i$-th attribute indicator, and each element $a_i^-$ of the negative ideal solution vector represents the worst value for the $i$-th attribute indicator.

**Step6**: Compute the distances between each alternative solution and the ideal solution. This can be done using the Euclidean distance or other distance metrics. The distance between an alternative solution and the ideal solution can be expressed as:

$$D_i^* = \sqrt{\sum_{j=1}^{m}(x_{ij}^* - a_j^*)^2} \quad (2)$$

**Step7**: Calculate the distances between each alternative solution and the negative ideal solution. This can be done using the Euclidean distance or other distance metrics. The distance between an alternative solution and the negative ideal solution can be expressed as:

$$D_i^- = \sqrt{\sum_{j=1}^{m}(x_{ij}^* - a_j^-)^2} \quad (3)$$

**Step8**: Based on the distances from the ideal solution and the negative ideal solution, calculate the comprehensive evaluation index for each alternative solution. This can be done using the following formula:

$$C_i = \frac{D_i^-}{D_i^+ + D_i^-} \quad (4)$$

Here, $D_i^+$ represents the distance between the alternative solution and the ideal solution.

**Step9**: Rank the alternative solutions based on the comprehensive evaluation index $C_i$ to assess their relative superiority.

Through the above steps, we have identified a model for the flow of the game-time game, which is specifically applied and visualized in the following.

## 3.2 Calculation by TOPSIS Model

We use TOPIS evaluation method, the number of discs, the number of games, the number of points scored, the serving side as evaluation indicators, the weights are 0.4, 0.25, 0.2, 0.15; select the first group of players in the game data, the player one and player two separate, respectively, record the two players of each moment of the score, the score of all moments visualization, the score of the player one and the player two scores are drawn as a line graph, the As shown in Figure 2.

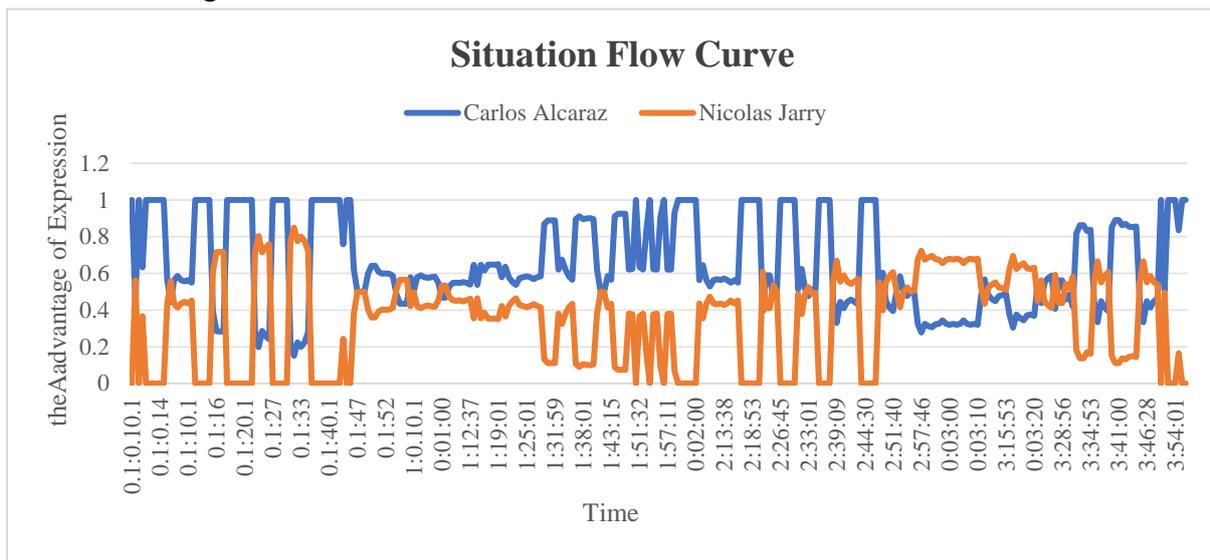

Figure 2: Scoring for Player 1 and Player 2

We can see from Figure 1 that player one and player two in each moment of the score, if at a certain moment player one score is higher than player two, it can indicate that at that moment player one's performance is better, the performance advantage is greater; due to the more moments, the figure in the point is denser, so first selected the first disk data, to get a clearer line graph as shown in Figure 3.

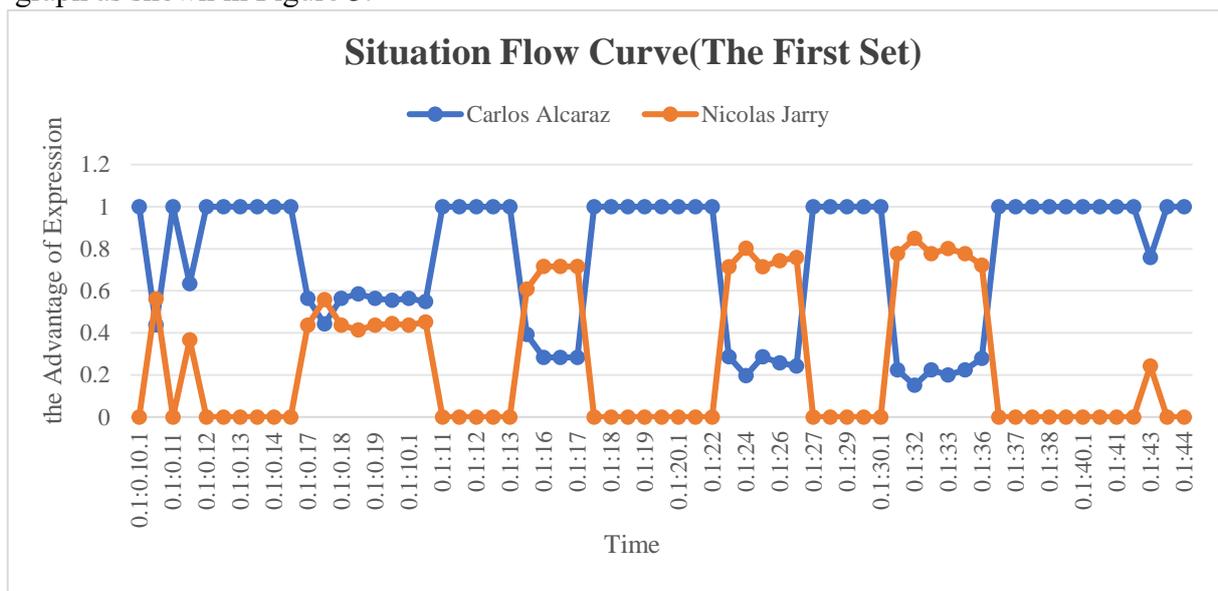

Figure 3: Scoring of the first set for Player 1 and Player 2

Similarly, the second disk of data was selected to plot a line graph.

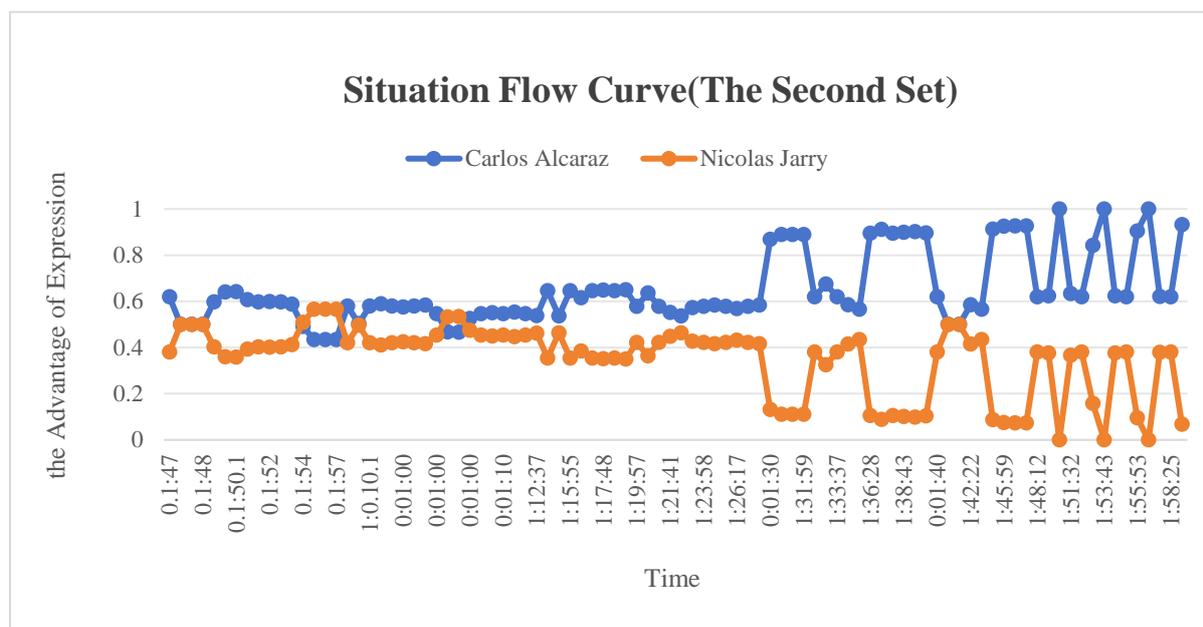

Figure 4: Scoring of the second set for Player 1 and Player 2

The performance and dominance of the two players at a given point in time can be seen in the above line graph. In the overall trend of the game, we can see that Player 1 always has the upper hand, but there is a more significant reversal by Player 2 in the moments from 2:54:43 to 3:28:56. Specifically in each set, we can see that in the first set Player 1 had a clear advantage, but in the three time periods from 1:16 to 1:17 Player 2 had a significant reversal; while in the second set Player 1 had a clear advantage, and Player 2 did not have an advantage for a long period of time. The above analysis also illustrates the existence of "momentum".

## 4 Analyzing the Randomness of "Momentum"

To solve this problem, we constructed a model using 0-1 logistic regression to determine whether "momentum" plays a role in the game.

### 4.1 Selection of 0-1 Logistic Regression Model

The 0-1 logistic regression is a machine learning algorithm used for binary classification problems. Its principle is based on the logistic function (also known as the sigmoid function) and maximum likelihood estimation.

Assuming we have a training dataset consisting of $m$ samples, each sample composed of $n$ features. Our goal is to predict the probability of each sample belonging to class 0 or class 1 based on these features.

Logistic regression utilizes the following hypothesis function to represent the probability of a sample belonging to class 1:

$$h_\theta(x) = \frac{1}{1+e^{-\theta^T x}} \tag{5}$$

Here, $x$ is a vector containing $n$ features, and $\theta$ is a vector containing $n$ parameters.

To determine the parameters $\theta$, we need to define a cost function that measures the difference between the predicted values and the actual values. For logistic regression, the commonly used cost function is the logistic loss function:

$$J(\theta) = -\frac{1}{m}\sum_{i=1}^{m}\left[y^{(i)}\log(h_\theta(x^{(i)})) + (1-y^{(i)})\log(1-h_\theta(x^{(i)}))\right] \quad (6)$$

Here, $y^{(i)}$ represents the actual class (0 or 1) of the $i$-th sample, and $x^{(i)}$ represents the feature vector of the $i$-th sample.

Our objective is to minimize the cost function $J(\theta)$ by adjusting the parameters $\theta$ to make the predicted values as close as possible to the actual values.

The most commonly used optimization algorithm is gradient descent. The idea behind gradient descent is to iteratively update the parameters $\theta$ to gradually reduce the cost function $J(\theta)$. The update rule is as follows:

$$\theta_j := \theta_j - \alpha \frac{\partial J(\theta)}{\partial \theta_j} \quad (7)$$

Here, $\alpha$ is the learning rate, which controls the step size of each update.

The specific implementation steps are as follows:
1. Initialize the parameters $\theta$, which can be randomly initialized or initialized with a zero vector.

2. Compute the predicted values $h_\theta(x)$ based on the current parameters $\theta$.

3. Calculate the cost function $J(\theta)$ based on the predicted values and the actual values.

4. Compute the partial derivative $\frac{\partial J(\theta)}{\partial \theta_j}$ of the cost function $J(\theta)$ with respect to the parameter $\theta_j$.

5. Update the parameters $\theta$ using gradient descent: $\theta_j := \theta_j - \alpha \frac{\partial J(\theta)}{\partial \theta_j}$.

6. Repeat steps 2-5 until a convergence criterion is met (e.g., the change in the cost function is below a certain threshold or the maximum number of iterations is reached).

## 4.2 Analysis of the randomness of "Momentum"

We used 0-1 logistic regression with player performance fluctuations as the independent variable and success in getting a point as the dependent variable. In the independent variables, we created dummy variables for *serve_width*, *serve_depth*, and *return_depth*. The data we selected were the various momentum indicators and scores of the two players in the first set of matches, and the results of the logistic regression for player one is shown in the following table.

**Table 2: Results of the Logistic Regression for Player 1**

|  | B | Standard Error | Wald | df | P-Value | Exp(B) |
|---|---|---|---|---|---|---|
| *p1_ace* | -39.795 | 15817.914 | 0.000 | 1 | 0.998 | 0.000 |
| *p1_winner* | 59.544 | 10321.119 | 0.000 | 1 | 0.995 | 7.237E+25 |
| *winner_shot_type* | -20.314 | 4269.261 | 0.000 | 1 | 0.996 | 0.000 |
| *p1_double_fault* | -0.854 | 17013.550 | 0.000 | 1 | 1.000 | 0.426 |
| *p1_unf_err* | -22.174 | 7746.428 | 0.000 | 1 | 0.998 | 0.000 |
| *p1_net_pt* | -21.292 | 13527.238 | 0.000 | 1 | 0.999 | 0.000 |
| *p1_net_pt_won* | 40.891 | 15736.502 | 0.000 | 1 | 0.998 | 5.735E+17 |
| *p1_break_pt* | -19.973 | 12044.151 | 0.000 | 1 | 0.999 | 0.000 |
| *p1_break_pt_won* | 41.341 | 21009.191 | 0.000 | 1 | 0.998 | 8.997E+17 |
| *p1_distance_run* | 0.020 | 0.013 | 2.291 | 1 | 0.130 | 1.020 |
| *speed_mph* | -0.006 | 0.011 | 0.337 | 1 | 0.562 | 0.994 |
| *serve_width* | 0.063 | 0.204 | 0.097 | 1 | 0.755 | 1.066 |
| *serve_depth* | 0.867 | 0.881 | 0.969 | 1 | 0.325 | 2.381 |
| *return_depth* | -0.725 | 0.309 | 5.522 | 1 | 0.019 | 0.484 |
| *constant* | 0.613 | 0.623 | 0.967 | 1 | 0.325 | 1.846 |

**Table 3: Predictions for Player 1**

| Measured Value | | Predicted Value | | Percentage of predictions correct |
|---|---|---|---|---|
| | | *point_victor* | | |
| | | 0 | 1 | |
| *point_victor* | 0 | 108 | 34 | 76.1 |
| | 1 | 21 | 136 | 86.6 |
| Overall Percentage | | | | 81.6 |

Note: Significance level is 0.05

For player 1, we can see that a score of 0 is predicted correctly 76.1% of the time, a score of 1 is predicted correctly 86.6% of the time, and the overall prediction is correct 81.6%.

Applying the same methodology, we can obtain the results of the player 2 logistic regression.

**Table 4: Results of the Logistic Regression for Player 2**

|  | B | Standard Error | Wald | df | P-Value | Exp(B) |
|---|---|---|---|---|---|---|
| *p2_ace* | -39.328 | 12909.786 | 0.000 | 1 | 0.998 | 0.000 |
| *p2_winner* | 59.124 | 8625.760 | 0.000 | 1 | 0.995 | 4.758E+25 |
| *winner_shot_type* | -20.452 | 3492.215 | 0.000 | 1 | 0.995 | 0.000 |
| *p2_double_fault* | -4.623 | 23594.899 | 0.000 | 1 | 1.000 | 0.010 |
| *p2_unf_err* | -21.162 | 5564.215 | 0.000 | 1 | 0.997 | 0.000 |

|  | B | Standard Error | Wald | df | P-Value | Exp(B) |
|---|---|---|---|---|---|---|
| *p2_net_pt* | -22.857 | 5994.277 | 0.000 | 1 | 0.997 | 0.000 |
| *p2_net_pt_won* | 42.172 | 8551.529 | 0.000 | 1 | 0.996 | 2.065E+18 |
| *p2_break_pt* | -19.915 | 15777.475 | 0.000 | 1 | 0.999 | 0.000 |
| *p2_break_pt_won* | 41.717 | 32432.552 | 0.000 | 1 | 0.999 | 1.310E+18 |
| *p2_distance_run* | -0.007 | 0.014 | 0.223 | 1 | 0.636 | 0.993 |
| *speed_mph* | -0.009 | 0.019 | 0.197 | 1 | 0.657 | 0.991 |
| *serve_width* | -0.330 | 0.364 | 0.822 | 1 | 0.365 | 0.719 |
| *serve_depth* | 3.492 | 1.745 | 4.002 | 1 | 0.045 | 32.840 |
| *return_depth* | -0.022 | 0.287 | 0.006 | 1 | 0.939 | 0.978 |
| *constant* | -0.386 | 0.641 | 0.363 | 1 | 0.547 | 0.680 |

Table 5: Predictions for Player 2

| Measured Value | | Predicted Value | | |
|---|---|---|---|---|
| | | *point_victor* | | Percentage of predictions correct |
| | | 0 | 1 | |
| *point_victor* | 0 | 144 | 14 | 91.1 |
| | 1 | 39 | 103 | 72.5 |
| Overall Percentage | | | | 82.3 |

Note: The cutoff value is 0.500

For player 2, we can see that a score of 0 is predicted correctly 91.1% of the time, a score of 1 is predicted correctly 72.5% of the time, and the overall prediction is correct 82.3%.

From the results of the logistic regression between Player 1 and Player 2, it is clear that the successive occurrences of a player's performance fluctuations and successes are not random, which means that the tennis coach's argument is not valid.

# 5 Indicators that Turn the Game Around

## 5.1 Developing Models to Predict Change

The situation change means that the game has come to the match point, so we choose the key inning data momentum indicator of each game as the indicator of situation change analysis, and because the characteristics of each person's hitting style are different, so we choose the data of the four games that player Carlos Alcaraz participated in to do the analysis, and we use the logistic regression method to predict the momentum change data of the fourth game with the actual occurrence when player Carlos Alcaraz is in an advantageous position. Using the logistic regression method, using the key inning momentum change of the first three games in which player Carlos Alcaraz participated as the sample data, and predicting the data of the momentum change of the fourth game with the actual occurrence of the logistic regression analysis, we can get the results of the logistic regression when the player Carlos Alcaraz is in an advantageous position, as shown in the following table.

**Table 6: Results of the Logistic Regression when Player Carlos Alcaraz is dominant**

|                      | B        | Standard Error | Wald  | df | P-Value | Exp(B)      |
|----------------------|----------|----------------|-------|----|---------|-------------|
| *p1_winner*          | 23.217   | 39403.507      | 0.000 | 1  | 1.000   | 1.211E+10   |
| *winner_shot_type*   | ----     | ----           | 0.000 | 2  | 1.000   | ----        |
| *winner_shot_type(1)*| 12.711   | 41886.869      | 0.000 | 1  | 1.000   | 331336.314  |
| *winner_shot_type(2)*| -136.275 | 52662.668      | 0.000 | 1  | 0.998   | 0.000       |
| *p1_double_fault*    | -102.306 | 113103.467     | 0.000 | 1  | 0.999   | 0.000       |
| *p1_unf_err*         | -1.504   | 60172.026      | 0.000 | 1  | 1.000   | 0.222       |
| *p1_net_pt*          | -102.362 | 72543.083      | 0.000 | 1  | 0.999   | 0.000       |
| *p1_net_pt_won*      | 144.905  | 84584.189      | 0.000 | 1  | 0.999   | 8.538E+62   |
| *p1_break_pt*        | -490.949 | 168767.507     | 0.000 | 1  | 0.998   | 0.000       |
| *p1_break_pt_won*    | 47.397   | 40738.748      | 0.000 | 1  | 0.999   | 3.841E+20   |
| *p1_distance_run*    | -1.135   | 2203.830       | 0.000 | 1  | 1.000   | 0.321       |
| *speed_mph*          | -3.480   | 1861.239       | 0.000 | 1  | 0.999   | 0.031       |
| *serve_4idth*        | -17.593  | 16970.767      | 0.000 | 1  | 0.999   | 0.000       |
| *return_1epth*       | -11.703  | 25763.787      | 0.000 | 1  | 1.000   | 0.000       |
| *constant*           | 494.751  | 140273.822     | 0.000 | 1  | 0.997   | 7.371E+214  |

**Table 7: Omnibus Test for Model Coefficients when Player Carlos Alcaraz is dominant**

| Chi-Square (math.) | df | Significance |
|--------------------|----|--------------|
| 37.393             | 13 | 0.000        |

**Table 8: Log-likelihood and R-square when Player Carlos Alcaraz is dominant**

| Log-likelihood | Cox Snell R-Square | Negolko R-square |
|----------------|--------------------|------------------|
| 0.000          | 0.750              | 1.000            |

Note: The cutoff value is 0.500

**Table 9: Predictions when Player Carlos Alcaraz is dominant**

| Measured Value |   | Predicted Value |    | Percentage of predictions correct |
|----------------|---|-----------------|----|-----------------------------------|
|                |   | *point_victor*  |    |                                   |
|                |   | 0               | 1  |                                   |
| *point_victor* | 0 | 13              | 0  | 100.0                             |
|                | 1 | 0               | 14 | 100.0                             |
| Overall Percentage |   |             |    | 100.0                             |

It can be obtained that the percentage of correctness can reach 100% and the prediction result is successful. This shows that there is a certain regularity in the momentum change indicator when the player Carlos Alcaraz is in an advantageous position in a key game, and the data of this indicator can be used to determine whether the player is in an advantageous position in a key game in a match; if so, the player remains in an advantageous position, and vice versa, the situation changes from one player to another.

Similarly, the results of the logistic regression analysis to find the player Carlos Alcaraz at a disadvantage can be obtained as shown in the following table.

Table 10: Results of the Logistic Regression when Player Carlos Alcaraz is inferior

|  | B | Standard Error | Wald | df | P-Value | Exp(B) |
|---|---|---|---|---|---|---|
| *p1_winner* | 62.548 | 41438.849 | 0.000 | 1 | 0.999 | 1.460E+27 |
| *winner_shot_type* | -21.133 | 16771.304 | 0.000 | 1 | 0.999 | 0.000 |
| *p1_unf_err* | -21.003 | 40192.970 | 0.000 | 1 | 1.000 | 0.000 |
| *p1_distance_run* | 0.095 | 0.080 | 1.404 | 1 | 0.236 | 1.100 |
| *speed_mph* | -0.291 | 679.634 | 0.000 | 1 | 1.000 | 0.748 |
| *serve_width* | 18.805 | 56722.396 | 0.000 | 1 | 1.000 | 1.468E+8 |
| *return_depth* | 0.769 | 1.003 | 0.587 | 1 | 0.443 | 2.157 |
| *constant* | -3.308 | 2.795 | 1.401 | 1 | 0.237 | 0.037 |

Table 11: Omnibus Test for Model Coefficients when Player Carlos Alcaraz is inferior

| Chi-Square (math.) | df | Significance |
|---|---|---|
| 13.183 | 7 | 0.068 |

Table 12: Log-likelihood and R-square when Player Carlos Alcaraz is inferior

| Log-likelihood | Cox Snell R-Square | Negolko R-square |
|---|---|---|
| 12.714 | 0.483 | 0.665 |

Note: The cutoff value is 0.500

Table 13: Predictions when Player Carlos Alcaraz is inferior

| Measured Value | | Predicted Value | | |
|---|---|---|---|---|
| | | *point_victor* | | Percentage of predictions correct |
| | | 0 | 1 | |
| *point_victor* | 0 | 12 | 1 | 92.3 |
| | 1 | 2 | 5 | 71.4 |
| Overall Percentage | | | | 85.0 |

It can be seen that the percentage of correctness can reach 85%, which is high and can indicate the success of the prediction. The indicator of the change in momentum of the player Carlos Alcaraz in the key game has a certain degree of regularity, which can be used in subsequent matches to determine whether or not he is in a disadvantageous position when the situation of the game changes, and if so, it can indicate a change in the dominant change from one player to the other.

## 5.2 Exploring the most Relevant Factors

The question we used the method of Spearman correlation coefficient analysis to correlate the indicators of momentum change with the bureau's score as shown in the figure below.

|  | ace | winner | winner_shot_type | double_fault | unf_err | net_pt | net_pt_won | break_pt | break_pt_wo | break_pt_mi | distance_run | speed_mph | serve_width | serve_depth | return_1epth | point_victor |
|---|---|---|---|---|---|---|---|---|---|---|---|---|---|---|---|---|
| ace | NaN | NaN | NaN | NaN | NaN | NaN | NaN | NaN | NaN | NaN | NaN | NaN | NaN | NaN | NaN | NaN |
| winner | NaN | 1 | 0.001386 | 0.747825 | 0.506731 | 0.452879 | 0.506731 | 0.948304 | 0.157117 | 0.325456 | 0.490369 | 0.62452 | 0.861063 | 0.971489 | 0.884519 | 0.007578 |
| winner_shot_type | NaN | 0.001386 | 1 | 0.552427 | 0.217934 | 0.163003 | 0.217934 | 0.548911 | 0.861323 | 0.571887 | 0.022719 | 0.122103 | 0.095408 | 0.126254 | 0.164258 | 0.247796 |
| double_fault | NaN | 0.747825 | 0.552427 | 1 | 0.000268 | 0.747825 | 0.77601 | 0.609363 | 0.807347 | 0.674142 | 0.174374 | 0.346626 | 0.345513 | 0.055541 | 0.249846 | 0.378406 |
| unf_err | NaN | 0.506731 | 0.217934 | 0.000268 | 1 | 0.286408 | 0.556975 | 0.84828 | 0.614823 | 0.58776 | 0.81274 | 0.938412 | 0.684081 | 0.248325 | 0.332749 | 0.06579 |
| net_pt | NaN | 0.452879 | 0.163003 | 0.747825 | 0.286408 | 1 | 3.13E-18 | 0.231197 | 0.569525 | 0.325456 | 0.987801 | 0.006203 | 0.000317 | 0.001986 | 0.005672 | 0.093686 |
| net_pt_won | NaN | 0.506731 | 0.217934 | 0.77601 | 0.556975 | 3.13E-18 | 1 | 0.29006 | 0.614823 | 0.38465 | 0.865651 | 0.008289 | 0.001384 | 0.018365 | 0.015223 | 0.019061 |
| break_pt | NaN | 0.948304 | 0.548911 | 0.609363 | 0.84828 | 0.231197 | 0.29006 | 1 | 0.000342 | 6.88E-14 | 0.357561 | 0.003845 | 0.003757 | 0.003027 | 0.011985 | 0.209754 |
| break_pt_won | NaN | 0.157117 | 0.861323 | 0.807347 | 0.614823 | 0.569525 | 0.614823 | 0.000342 | 1 | 0.456629 | 0.522629 | 0.181387 | 0.180371 | 0.171138 | 0.241594 | 0.046119 |
| break_pt_missed | NaN | 0.325456 | 0.571887 | 0.674142 | 0.58776 | 0.325456 | 0.38465 | 6.88E-14 | 0.456629 | 1 | 0.531376 | 0.018958 | 0.018671 | 0.016176 | 0.04069 | 0.005493 |
| distance_run | NaN | 0.490369 | 0.022719 | 0.174374 | 0.81274 | 0.987801 | 0.865651 | 0.357561 | 0.522629 | 0.531376 | 1 | 0.418551 | 0.247463 | 0.367644 | 0.836738 | 0.530343 |
| speed_mph | NaN | 0.62452 | 0.122103 | 0.346626 | 0.938412 | 0.006203 | 0.008289 | 0.003845 | 0.181387 | 0.018958 | 0.418551 | 1 | 1.22E-24 | 6.13E-26 | 9.46E-15 | 0.001401 |
| serve_width | NaN | 0.861063 | 0.095408 | 0.345513 | 0.684081 | 0.000317 | 0.001384 | 0.003757 | 0.180371 | 0.018671 | 0.247463 | 1.22E-24 | 1 | 3.03E-28 | 5.93E-15 | 0.00161 |
| serve_depth | NaN | 0.971489 | 0.126254 | 0.055541 | 0.248325 | 0.001986 | 0.018365 | 0.003027 | 0.171138 | 0.016176 | 0.367644 | 6.13E-26 | 3.03E-28 | 1 | 5.89E-16 | 0.006796 |
| return_1epth | NaN | 0.884519 | 0.164258 | 0.249846 | 0.332749 | 0.005672 | 0.015223 | 0.011985 | 0.241594 | 0.04069 | 0.836738 | 9.46E-15 | 5.93E-15 | 5.89E-16 | 1 | 0.008607 |
| point_victor | NaN | 0.007578 | 0.247796 | 0.378406 | 0.06579 | 0.093686 | 0.019061 | 0.209754 | 0.046119 | 0.005493 | 0.530343 | 0.001401 | 0.00161 | 0.006796 | 0.008607 | 1 |

**Figure 5: Heat map for Spearman's correlation coefficient analysis**

The P-matrix of the above correlation analysis shows the correlation of each variable with point-victor, and it can be concluded that *point_victory* has the strongest correlation with the winner's variables *net_pt_won, break_pt_won, break_pt_missed, speed_mph, serve_width, serve_depth*, and *return_1epth*.

Also, because the correlation coefficients of these eight variables in the P-matrix are not very different, in order to get the most correlated factors, these eight variables were analyzed by principal component analysis as shown in the figure below.

|  | F1 | F2 | F3 | F4 | F5 | F6 | F7 | F8 |
|---|---|---|---|---|---|---|---|---|
| p1_winner | 0 | 0.64 | -0.35 | 0.61 | 0.3 | 0.02 | -0.07 | -0.04 |
| p1_net_pt_won | -0.25 | -0.14 | 0.76 | 0.38 | 0.36 | 0.13 | -0.2 | -0.05 |
| p1_break_pt_won | 0.11 | 0.6 | 0.47 | -0.05 | -0.64 | -0.01 | 0 | 0.02 |
| p1_break_pt_missed | 0.21 | -0.46 | -0.12 | 0.68 | -0.53 | 0 | -0.04 | 0.02 |
| speed_mph | -0.49 | 0.01 | -0.12 | 0 | -0.13 | 0.05 | -0.25 | 0.81 |
| serve_width | -0.46 | 0 | 0.06 | 0.14 | -0.05 | -0.63 | 0.6 | -0.05 |
| serve_depth | -0.47 | -0.01 | -0.17 | -0.08 | -0.22 | -0.22 | -0.61 | -0.52 |
| return_depth | 0.45 | -0.01 | 0.11 | -0.02 | 0.18 | -0.73 | -0.39 | 0.25 |
|  |  |  |  |  |  |  |  |  |
| Contribution Rate | 0.48 | 0.16 | 0.11 | 0.1 | 0.09 | 0.03 | 0.02 | 0.01 |
| Cumulative Contribution | 0.48 | 0.64 | 0.76 | 0.85 | 0.94 | 0.97 | 0.99 | 1 |

**Figure 6: Principal Component Analysis Heat Map**

The cumulative contribution of F1, F2, and F3 amounted to 75%, and F1, F2, and F3 were taken as principal components. The first principal component F1 has moderate positive loadings on *speed_mph, serve_width, serve_depth,* and moderate negative loadings on *return_depth*, so we call the first principal component the player's subjective adjustable component.

The second principal component F2 has medium positive loadings on *p1_winner, p1_break_pt_won* and medium negative loadings on *p1_break_pt_missed*, and the third principal component F3 has large positive loadings on *p1_net_pt_won* and medium negative loadings on *p1_break_pt_won*. The second and third principal components are the components of the match in which objective and subjective factors jointly regulate the results.

Among them, the first principal component has a high contribution of 48%, and we can show that *speed_mph, serve_width*, and *serve_depth* are the most relevant factors to *point_victor*.

## 5.3 Exploring the most Relevant Factors

Given the differences in "momentum" changes in past matches, we recommend that players analyze their past records against specific opponents to understand their weaknesses and strengths before the match. This will help the player to develop a more effective strategy. Maintaining good physical condition to prevent injury and fatigue will help to maintain a consistently high level of performance during matches. Develop strong mental attributes, including the ability to handle pressure, stay focused and cope with challenges.

# 6 Accuracy Testing of Model

## 6.1 Build the model and analyze the results

The model is applied to other matches by selecting the match data of player Alexander Zverev and player Matteo Berrettini from the second match of the given data.

First of all, the performance of the two players in that game is analyzed by doing TOPSIS evaluation class model, and the scores of the two players at each moment are visualized to get the line graph.

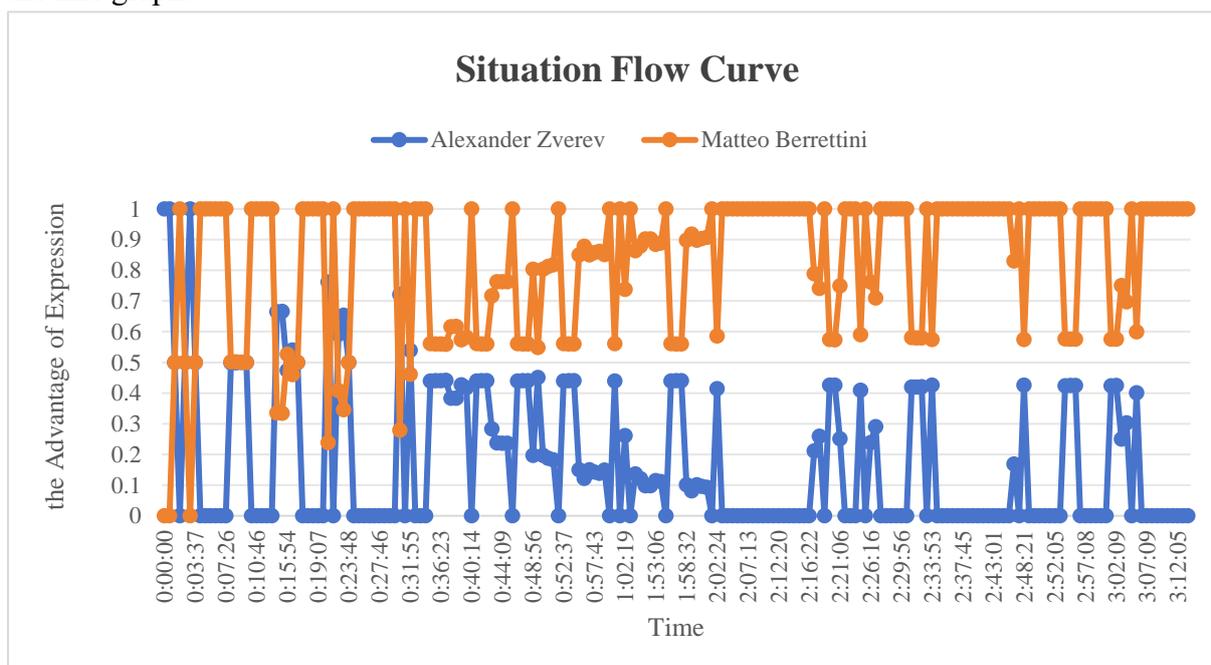

**Figure 7: Points scored by both players at various moments**

From the line graph can be seen in the game at various moments of the performance of the two players and the advantages, can be seen in the middle and second half of the game player 2's performance is better than the performance of player 1; due to the first two discs of the player's performance fluctuations and points are denser, it is not easy to see, the first two discs of the data scores were organized to do the visualization of the following chart.

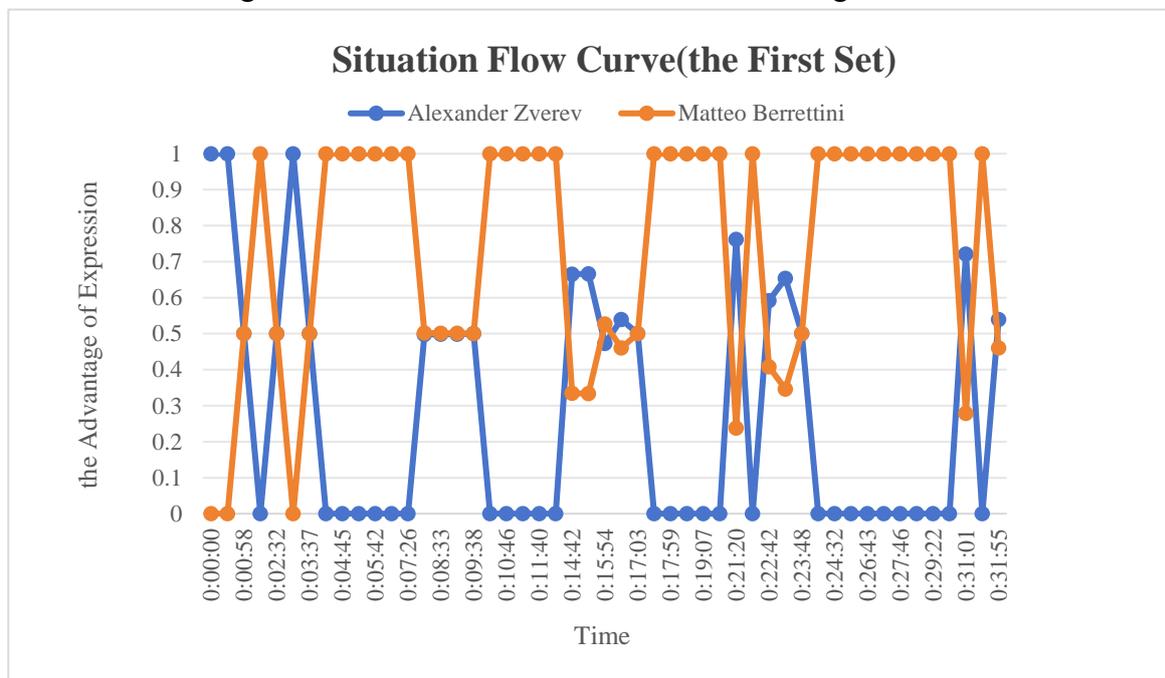

**Figure 8: Points scored by both players at various moments in the first set**

From the line graph, it can be seen that the performance and advantage of the two players fluctuated a lot at the beginning of the first set, and after 0:03:37 the performance and advantage of player 2 came to the forefront, and although there were still fluctuations, it was still clear that the performance and advantage of player 2 was better than that of player 1.

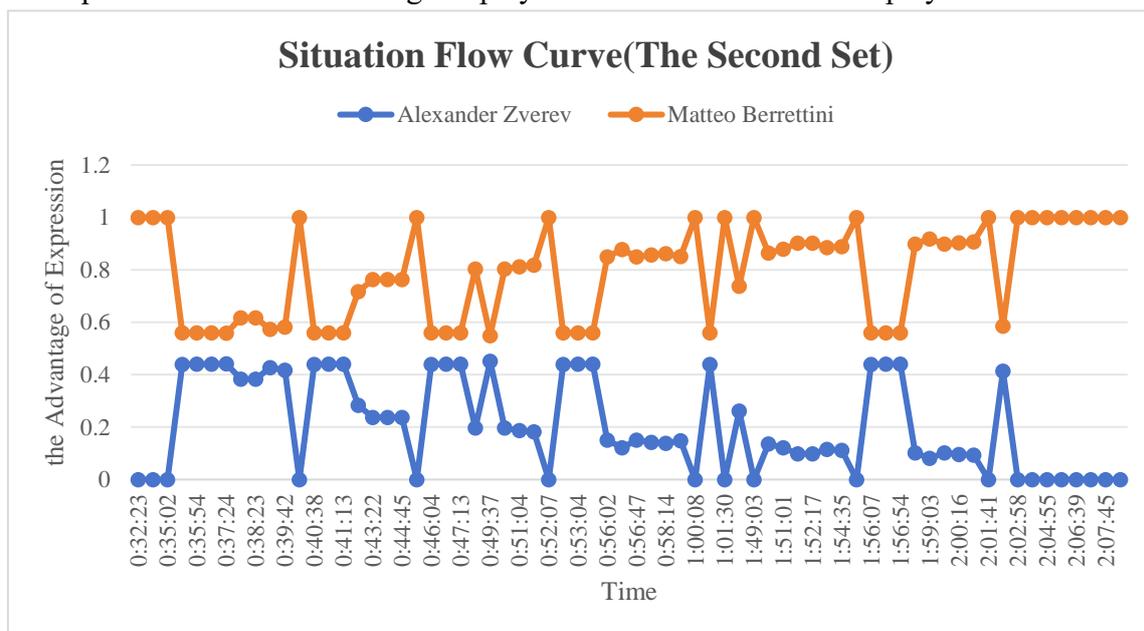

**Figure 9: Points scored by both players at various moments in the second set**

As we can see from the graph above, player 2 maintained his advantage at the start of the second set and continued to lead player 1.

Situation change means that the game comes to the match point, so the key inning data momentum indicator of each game is chosen as the indicator of situation change analysis, and because the characteristics of each person's playing style are different, so the data of the four games in which the player participated are chosen to be analyzed. We use the logistic regression method to take the key inning momentum change of the first two games in which Matteo Berrettini participated as sample data, and predict the key inning momentum change of the last three games of the second game with the actual occurrence of logistic regression analysis, and we can get the results of logistic regression when Matteo Berrettini is in the advantage, which are as shown in the following table.

Table 14: Results of the Logistic Regression for Player 2

|  | B | Standard Error | Wald | df | P-Value | Exp(B) |
|---|---|---|---|---|---|---|
| p2_ace | -42.579 | 59255.745 | 0.000 | 1 | 0.999 | 0.000 |
| p2_winner | 35.455 | 43540.418 | 0.000 | 1 | 0.999 | 2.500E+18 |
| winner_shot_type | -15.203 | 40192.970 | 0.000 | 1 | 1.000 | 0.000 |
| p2_unf_err | -18.024 | 40192.970 | 0.000 | 1 | 1.000 | 0.000 |
| p2_net_pt | 27.618 | 13807.783 | 0.000 | 1 | 0.998 | 9.868E+11 |
| p2_distance_run | -0.636 | 0.425 | 2.238 | 1 | 0.135 | 0.530 |
| speed_mph | -0.280 | 0.228 | 1.506 | 1 | 0.220 | 0.756 |
| serve_width | 1.756 | 1.851 | 0.900 | 1 | 0.343 | 5.790 |
| serve_depth | 18.724 | 21.027 | 0.793 | 1 | 0.373 | 1.354E+8 |
| return_depth | -4.395 | 6.107 | 0.518 | 1 | 0.472 | 0.012 |
| constant | 19.427 | 16.862 | 1.327 | 1 | 0.249 | 2.737E+8 |

Table 15: Predictions when Player Carlos Alcaraz is inferior

| Measured Value | | Predicted Value | | Percentage of predictions correct |
|---|---|---|---|---|
| | | point_victor | | |
| | | 0 | 1 | |
| point_victor | 0 | 4 | 2 | 66.7 |
| | 1 | 2 | 12 | 85.7 |
| Overall Percentage | | | | 80.0 |

The percentage of correctness can reach 80% and the prediction result is successful. It shows that there is a certain regularity in the momentum change indicator when the player Matteo Berrettini is in the situation change in the key game, which can be based on the data of this indicator to determine whether the player is in a dominant position in the key game in the game, if yes, the player still maintains the dominant position, and vice versa, the situation change from one player to another.

To summarize, our model is more accurate in predicting the change of "Momentum" in the game, and can be used for practical development.

## 6.2 Factors to be Included in Future Models

We constructed a model to analyze the situation of tennis matches using TOPSI and logistic regression methods, and the percentage of successful predictions could reach about 80% when applying the model to other matches. However, the error may be due to the fact that the changes in the "momentum" during the match are influenced by the interplay of various factors.

In order to further improve the accuracy and applicability of the model, we believe that we can consider adding the psychological quality of the players, the physical quality of the players, and the analysis of the opponent's strengths and weaknesses to improve the model and increase the universality and correctness of the model.

1. the psychological quality of the players: Psychological quality plays an important role in the game. We can add players' psychological quality as a factor in the model, for example, players' self-confidence, concentration, ability to cope with pressure and so on. These factors may affect a player's performance in a match, so including them in the model may help improve prediction accuracy.

2. Players' physical fitness: Players' physical fitness is also an important factor that affects the outcome of a match. We can use players' physical quality indicators as one of the characteristics, for example, height, weight, speed, strength, and so on. These indicators may have an impact on a player's performance, so taking them into account may help improve the accuracy of the model.

3. Analysis of opponents' strengths and weaknesses: The strengths and weaknesses of opponents can have a significant impact on the outcome of a match. We can incorporate relevant information about the opponent into the model, such as the opponent's ranking, historical performance, technical characteristics, etc. By analyzing the opponent's strengths and weaknesses, your model can better predict the outcome of the match.

4. Match venue: The match venue may also have an impact on the match result. We can add the characteristics of the playing field as a feature in the model, such as field type (grass, hard surface, red soil, etc.), altitude, temperature, etc. These factors may affect player performance and the outcome of the match, so taking them into account may help improve the accuracy of the model.

By adding factors such as players' mental fitness, physical fitness, analysis of opponents' strengths and weaknesses characteristics, and playing venues, we can refine the model and increase its generalizability and correctness.

## 6.3 Generalizability of the Model to Other Competitions

Our model mainly utilizes **TOPSIS and Logistic Regression Model**, where the variables are commonly used in tennis tournaments. Our model is trained using only the variables and data from tennis matches, rather than being applicable only to one type of match, the tennis match.

By analyzing the data, our model can be well extended to many sports competitions with different genders, different types, and different scoring mechanisms. As long as we get the variables and data of the relevant game, we can use our model to study the change of "momentum" in the sports game. For example, in a basketball game, as long as we specify the number of two-pointers, three-pointers, rebounds, and other data between two players, we can construct a TOPSIS model that depicts the flow of the situation, and we can also use logistic regression models to analyze the "momentum" of the game, which can be used to give the players and coaches advice on the court.

# 7 Model Evaluation and Further Discussion

## 7.1 Strengths

- The TOPSIS method provides a quantitative analysis of the metrics, while 0-1 logistic regression provides a qualitative analysis of the dichotomous problem. By combining these two methods, we can synthesize player data and performance to more fully assess the match situation.

- Both TOPSIS and 0-1 logistic regression are widely used and validated methods that have achieved good predictive results in many fields. Therefore, models constructed using these two methods may have a high success rate in predictive analysis.

- By correlating the force or momentum gained by a player through a movement or series of movements with the scoring and situational changes of a game, our model provides an intuitive visualization to help analyze situational changes.

- Our model allows us to see how the situation changes and the performance advantages between players by analyzing the match data of both players. This allows us to better understand the dynamics of the game and make predictions and decisions accordingly.

## 7.2 Weaknesses

- Our model uses a combination of TOPSIS and 0-1 logistic regression, which requires high data quality. Accurate, comprehensive, and reliable player data and match data, as well as corresponding labels (e.g., match results, situation changes, etc.) are required. If the data quality is low or lacks key information, the accuracy and reliability of the model may be affected.

- Our model ignores factors such as players' physical and psychological qualities, playing surfaces, players' playing styles and bat speeds, and exploitation of opponents' weaknesses, making the model slightly less comprehensive.

## 7.3 Further Discussion

- We will incorporate more factors into our model so that our model can explore more variables and solve more problems.

- We will use more data from the tournament to improve our model, which will increase the

generalizability of our model.

- By reading more literature, we can extend this model to more areas of sports and make it more useful.

# 8 Conclusion

In our paper, we have designed **two models** that are closely related to each other, both of which can be used to determine the "momentum" of a game and to predict the future of the game from information about the past of the game. The first model is the **TOPSIS Model**, which we chose after comparing several algorithms, and which helps us to explain the existence of "momentum". The second model is a **0-1 Logistic Regression Model**, which we programmatically solved and predicted the future of the game. At the same time, we explored other independent variables that might be incorporated into our model and that might enrich our model. However, due to data limitations, there is still a large gap between our model and reality. In the future, we will continue to improve our model in order to better explain the significance of "momentum" in sports.